\documentclass[reqno,11pt]{amsproc}
\usepackage{amssymb}
\usepackage{amsmath}
\usepackage{amsfonts}
\usepackage{microtype}
\usepackage{xcolor}
\usepackage{geometry}
\usepackage{mathrsfs}
\usepackage[color=white,textcolor=red]{todonotes}

\usepackage[natbib,style=authoryear,backref=true,backend=bibtex]{biblatex}
\usepackage{tikz,tikz-cd}
\usetikzlibrary{arrows,cd,decorations.markings,arrows.meta}
\tikzset{->-/.style={decoration={markings,mark=at position .5 with {\arrow{>}}},postaction={decorate}}}
\tikzset{-<-/.style={decoration={markings,mark=at position .5 with {\arrow{<}}},postaction={decorate}}}
\tikzstyle{box}=[fill=white, draw=black, shape=rectangle, inner sep=14pt]
\tikzstyle{forward arrow}=[->-]
\tikzstyle{backward arrow}=[-<-]

\definecolor{myurlcolor}{rgb}{0,0,0.4}
\definecolor{mycitecolor}{rgb}{0,0.5,0}
\definecolor{myrefcolor}{rgb}{0.5,0,0}
\usepackage{hyperref}
\hypersetup{colorlinks,
linkcolor=myrefcolor,
citecolor=mycitecolor,
urlcolor=myurlcolor}
\usepackage[capitalize]{cleveref}

\newcommand{\beq}{\begin{equation}}
\newcommand{\eeq}{\end{equation}}
\newcommand{\N}{\mathbb{N}}

\newcommand{\R}{\mathbb{R}}
\newcommand{\C}{\mathbb{C}}
\newcommand{\op}{\mathrm{op}}
\newcommand{\id}{\mathsf{id}}		% identity morphism
\newcommand{\cat}[1]{\mathsf{#1}}	% font for categories
\newcommand{\Set}{\cat{Set}}		% category of sets
\newcommand{\Mfd}{\cat{Mfd}}		% the category of manifolds
\newcommand{\ST}{\mathsf{Spacetimes}}	% category of manifolds wrt which a field can be pulled back
\newcommand{\LD}{\mathcal{L}}		% Lagrangian density
\newcommand{\ff}{\mathcal{F}}		% field functor
\newcommand{\fa}{\mathcal{A}}		% \ff-algebra
\newcommand{\df}{\mathcal{D}}		% density functor
\theoremstyle{plain}
\newtheorem{dummy}{Dummy}[section]
\Crefname{thm}{Theorem}{Theorems}
\Crefname{lem}{Lemma}{Lemmas}
\Crefname{prop}{Proposition}{Propositions}
\Crefname{cor}{Corollary}{Corollaries}
\Crefname{conj}{Conjecture}{Conjectures}
\Crefname{qstn}{Question}{Questions}
\newtheorem{defn}[dummy]{Definition}\Crefname{defn}{Definition}{Definitions}
\Crefname{nota}{Notation}{Notations}
\Crefname{prob}{Problem}{Problems}
\theoremstyle{remark}
\Crefname{ex}{Example}{Examples}
\newtheorem{rem}[dummy]{Remark}\Crefname{rem}{Remark}{Remarks}
\Crefname{note}{Note}{Notes}
\numberwithin{equation}{section}
\Crefname{equation}{}{}		% abbreviate 'Equation (3.2)' to '(3.2)'

\newtheorem{scalar_ex}[dummy]{Scalar Field}\Crefname{scalar_ex}{Scalar Field}{Scalar Field}
\newtheorem{electro_ex}[dummy]{Electromagnetic Field}\Crefname{electro_ex}{Electromagnetic Field}{Electromagnetic Field}
\newtheorem{spinor_ex}[dummy]{Spinor Field}\Crefname{spinor_electro_ex}{Spinor Field}{Spinor Field}
\newtheorem{spinor_electro_ex}[dummy]{Spinor Electrodynamics}\Crefname{spinor_electro_ex}{Spinor Electrodynamics}{Spinor Electrodynamics}
\newtheorem{gr_ex}[dummy]{Metric Field}\Crefname{gr_ex}{Metric Field}{Metric Field}
\newtheorem{scalar_fun_ex}[dummy]{Scalar Field (Kinematic)}\Crefname{scalar_fun_ex}{Scalar Field (Kinematic)}{Scalar Field (Kinematic)}

\allowdisplaybreaks

\let\originalleft\left
\let\originalright\right
\renewcommand{\left}{\mathopen{}\mathclose\bgroup\originalleft}
\renewcommand{\right}{\aftergroup\egroup\originalright}

\usepackage{enumitem}
\setlist[enumerate]{label=(\alph*),itemsep=5pt,topsep=8pt}
\setlist[itemize]{label=$\triangleright$,itemsep=5pt,topsep=6pt}

\Crefformat{enumi}{#2#1#3}

\begin{document}

%-------------------------------------------------------------------

%%%%%%%%%%%% title page stuff %%%%%%%%%%%%%%%%%%%%%%%%%%

\title[An algebraic approach to physical fields]{An Algebraic Approach to Physical Fields}

\author{Lu Chen}
\address{Department of Philosophy, Ko{\c{c}} University, Istanbul, Turkey}
\email{luchen@ku.edu.tr}

\author{Tobias Fritz}
\address{Department of Mathematics, University of Innsbruck, Austria}
\email{tobias.fritz@uibk.ac.at}


\thanks{\textit{Acknowledgments.} We thank the anonymous referees at  \emph{Studies in History and Philosophy of Science} for their helpful suggestions on improving the paper.}

\begin{abstract}
	According to the algebraic approach to spacetime, a thoroughgoing dynamicism, physical fields exist without an underlying manifold. This view is usually implemented by postulating an algebraic structure (e.g., commutative ring) of scalar-valued functions, which can be interpreted as representing a scalar field, and deriving other structures from it. In this work, we point out that this leads to the unjustified primacy of an undetermined scalar field. Instead, we propose to consider algebraic structures in which all (and only) physical fields are primitive. We explain how the theory of \emph{natural operations} in differential geometry---the modern formalism behind classifying diffeomorphism-invariant constructions---can be used to obtain concrete implementations of this idea for any given collection of fields.

	    For concrete examples, we illustrate how our approach applies to a number of particular physical fields, including electrodynamics coupled to a Weyl spinor.
\end{abstract}

\maketitle

\tableofcontents

\section{Spacetime algebraicism}

Is spacetime a substance? One reason for answering yes is that the curvature of spacetime can explain phenomena such as light bending near massive objects according to general relativity.
The view that spacetime is a substance independent of the things and processes in it is called \textit{(spacetime) substantivalism}. The structure of spacetime is standardly represented by a smooth manifold equipped with a metric in manifold-based differential geometry.  But there are also many reasons for thinking that spacetime does not exist at the fundamental level.  For example, \citet{geroch}, among many others, has pointed out that a quantum theory of gravity calls for a ``smearing out'' of spacetime points. There is also a famous argument against substantivalism known as the hole argument, which purports to show that substantivalism is incompatible with even the weakest form of determinism~\citep{EN}. As a result, \citet{geroch} and~\citet{earman} respectively suggest \textit{(spacetime) algebraicism}, the view that physical fields exist fundamentally without an underlying spacetime, as a way to circumvent these difficulties. A similar approach has been taken up by the many works investigating noncommutative approaches to spacetime~\citep{DFR,marcolli}.

Here's very roughly how the standard version of algebraicism works. Consider the smooth manifold $M$ that represents our spacetime under substantivalism. Take all the smooth (real-valued or complex-valued) functions on $M$. We can operate on these functions in various ways. For example, we can add or multiply two smooth functions to get another. Now, instead of considering these smooth functions as maps $M \to \R$, we can instead consider them abstractly as elements of an algebraic structure defined by algebraic operations such as addition and multiplication.
The observation which launches algebraicism is that a large amount of information about the manifold can be encoded in this algebra of functions~\citep{connes_reconstruction}. In particular, all geometric entities that we need to do physics up to general relativity, including vectors and tensors, can be defined in algebraic terms without any reference to manifolds~\citep{geroch,pessers}. 
Geroch's proposal amounts to extending such a structure by an algebraic equivalent of the metric, resulting in an ``Einstein algebra'' whenever the Einstein field equations hold.

However, there is also bad news for algebraicism. \citet{rynasiewicz} pointed out that the approach of Einstein algebras neither addresses the concern from quantum gravity nor the hole argument. It does not necessarily smear out spacetime points, because points can be reconstructed from the algebraic structure.\footnote{Note that substantivalism is the view that spacetime exists at the fundamental level, which should be distinguished from the position that spacetime points can be constructed from the fundamental structures. It is also worth mentioning that Rynasiewicz's conclusion is contended by~\citet{bain}.} It does not avoid the hole argument because there is an analogous indeterminism among isomorphic algebras. 

So why should we still care about algebraicism? First, the advantages of Einstein algebras as an alternative approach to spacetime are still underexplored. For example, \citet{menon} connects Einstein algebras with the \textit{dynamic approach} to relativistic theories, according to which the geometry of spacetime is not fundamental and should be understood in terms of dynamic fields (rather than the other way around). The dynamic approach is famously advocated by Harvey Brown among others \citep{brown}. However, the traditional dynamic approach, though it disregards the metric as fundamental, still takes the metrically amorphous manifold as ontologically prior to physical fields. Algebraicism is a more thoroughgoing dynamic approach by getting rid of the underlying manifold---arguably an improvement upon this perfectly respectful approach to physics. Second, algebraicism is much more than the formalism of Einstein algebras, which is merely one implementation of algebraicism. For example, we could have algebraic structures that do not correspond to any geometric structures with points, such as the noncommutative algebras on which noncommutative geometry builds. Such algebraic structures may be helpful in formalizing quantum theories; see e.g.~\citet{bain,HS}.
Finally, since algebraicism is claimed to be an alternative to (manifold-theoretic) substantivalism as a foundation that is at least as good, it would be too dogmatic to subscribe to substantivalism without giving due consideration to algebraicism.

In this paper, we undertake some foundational work, both conceptually and technically, in exploring the potential of algebraicism as a dynamic approach. In order to take algebraicism realistically, we need to interpret what things fundamentally exist according to algebraicism. In the case of Einstein algebras, the fundamental structure is the algebra of all smooth functions $C^\infty(M)$ on the manifold $M$ that standardly represents our spacetime.  Those smooth functions can be understood as the possible configurations of a scalar field, which constitutes the fundamental ontology of Einstein algebras \citep{earman,DHL,bain}.
But what is this scalar field? Is it an actual physical field like the Higgs field?\footnote{Technically the answer can be seen to be ``no'', since the Higgs field is $\C^2$-valued rather than merely real-valued or complex-valued. A better candidate would be the hypothetical \emph{inflaton field}, which is indeed thought to be a real-valued scalar field. However, for the purposes of this discussion it does not matter which physical field it is, so let us pretend for the sake of discussion that it could be the Higgs field.} Here we seem to face a dilemma. On the one hand, if it is a physical field such as the Higgs, then we would be privileging one type of physical field ontologically, which seems arbitrary.
In particular, following the prescription that all relevant structures are to be defined in terms of the algebra $C^\infty(M)$, non-scalar fields such as the electromagnetic field then would have to be defined in terms of the Higgs field, but it is implausible to think that the electromagnetic field is ontologically dependent on the Higgs field.
On the other hand, if the elements of $C^\infty (M)$ are not physical fields, then taking them as fundamental objects amounts to positing fundamental ghost fields that play no role in our best current physical theory.  This would violate the empiricist spirit that largely motivates the dynamic approach to physics.\footnote{This would be less of an issue for substantivalists. When \citet{EN} consider algebraicism as a potential solution to the hole argument, they construe it as ``algebraic substantivalism.'' There, the idea is that the algebraic structure is a substitute for the spacetime manifold without necessarily being a physical field of its own. We thank the referee for bringing this point to our attention.} 

To avoid this dilemma, we need to go beyond the formalism of algebras of scalar-valued functions. The fundamental fields featured in our best physical theory should be ontologically on a par and independent from each other. This posits a nontrivial technical challenge. But it seems to be a necessary step towards building a conceptually perspicuous theory rather than one merely technically adequate.

To alleviate this problem, or at least to make some progress on it, we  will propose algebraic structures for fundamental physical fields in a way that does not privilege any of them over others.
To do so, we will propose a principled strategy for systematically determining fundamental algebraic structures based on the physical field content (although we will not argue that it is the only strategy). The hope is that the resulting dynamical theory is as perspicuous and as parsimonious as possible.

From a slightly different perspective, we will advocate the following conceptual shift.
Traditionally, physical fields are described by structures defined in terms of some primitive ones. In the standard algebraic approach, the algebra of scalar-valued functions is the primitive structure, with all other objects such as non-scalar fields described in terms of non-primitive derived structures.\footnote{This is analogous to the manifold approach, where the smooth manifold is the primitive structure and all other structures are defined on it.} Schematically, this situation can be depicted like this:
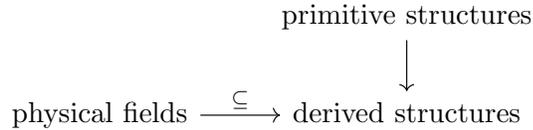
\[
	\begin{tikzcd}
									& \text{primitive structures} \ar{d}	\\
		\text{physical fields} \ar{r}{\subseteq}	& \text{derived structures}
	\end{tikzcd}
\]
 Our goal is to achieve the following picture instead:
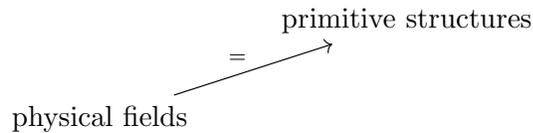
\[
	\begin{tikzcd}
									& \text{primitive structures} 	\\
		\text{physical fields} \ar{ur}{=}	
	\end{tikzcd}
\]
Like this, there is no fundamental structure underlying physical fields, which marks a thorough break from the tradition of positing spacetime or its surrogates such as a scalar field (which may not have physical reality). Under this approach, we can confidently affirm that physics can indeed be written ``on thin air alone" without ``the support of various space-time structures" \citep{earman}. 

Also, the need for considering any derived structures disappears: to do physics, it is enough to work with the primitive structures only. This makes the task of a realistic interpretation easy: we no longer need to worry about interpreting the derived structures, which are not limited to physical fields (for example, see \citet{AD}). Relatedly, the new framework is potentially ontologically more parsimonious than the standard algebraic approach in that there are fewer pieces of structure to quantify over. 

As a cost, the new framework is less parsimonious in its ideology since all physical fields are described by primitive structures. Such a concession is commonplace in theory choice, and is generally considered worthwhile if the resulting theory is more ontologically perspicuous (for example, consider the nominalist projects about numbers).  Of course, it still is a cost, and we aim to keep this cost minimal by only positing primitive structures that seem necessary, natural and joint-carving in light of our best physical theories. 

\section{Spacetime algebraicism based on a scalar field}

In this section, we provide some background by outlining our perspective on the algebraic approach to spacetime proposed by \citet{geroch,heller}, in which spacetime is characterized by its algebra of scalar-valued functions without explicit reference to spacetime points. While similar ideas also feature prominently in noncommutative geometry and its applications to physics~\citep{connes,marcolli}, our emphasis here is motivated by the closer proximity of Geroch's Einstein algebras to our proposal for non-scalar fields presented in the following sections.

As alluded to previously, the basic idea is to start with $C^\infty(M)$, the set of real-valued smooth functions on a manifold $M$, and to carefully analyze the following two questions:
\begin{enumerate}[label=(\arabic*)]
	\item\label{which_algebraic} Which algebraic structures does $C^\infty(M)$ carry at all?
	\item\label{which_physics} Which of these structures on $C^\infty(M)$ are relevant for formulating the kind of physics we want to consider?
\end{enumerate}
Both of these questions are important, and the rest of this paper will present some considerations on how to approach similar questions for non-scalar fields. Here we discuss them in the scalar field case, with Geroch's intended application to general relativity in mind.

Starting with \ref{which_algebraic}, we had already alluded to the fact that scalar-valued functions $M \to \R$ can be added and multiplied. They also possess an operation of scalar multiplication by real numbers. Taken together, these operations make $C^\infty(M)$ into a \emph{commutative algebra over $\R$}, and it is therefore natural to expect this to constitute a viable answer to \ref{which_algebraic}. This is indeed the approach adopted by Geroch and Heller, as well as the starting point of noncommutative geometry\footnote{With two technically relevant caveats:
	\begin{enumerate}[label=(\roman*)]
		\item One usually considers complex-valued scalar functions rather than real-valued ones.
		\item The algebras of smooth functions in noncommutative geometry are typically equipped with additional \emph{analytic} structure, again modelled after that of $C^\infty(M)$.
	\end{enumerate}
	Neither of these differences affects our conceptual discussion. We will not entertain analytic structures further in this paper, so suffice it to mention that this structure in particular serves as a stand-in for the additional algebraic structure considered in the next paragraph.}. Note that defining these operations does not require the introduction of coordinates on $M$, and therefore they are diffeomorphism-invariant.

However, this does not yet need to be the final answer, as there is some additional algebraic structure on $C^\infty(M)$ not captured by it being a commutative algebra over $\R$. Indeed if $g : \R \to \R$ is any smooth function, then we have an additional unary algebraic operation given by
\[
	C^\infty(M) \longrightarrow C^\infty(M),
	\qquad f \longmapsto g(f).
\]
More generally, every $g : \R^n \to \R$ defines an $n$-ary operation given by
\[
	\underbrace{C^\infty(M) \times \ldots \times C^\infty(M)}_{n \textrm{ factors}} \longrightarrow C^\infty(M),
	\qquad (f_1,\ldots,f_n) \longmapsto g(f_1,\ldots,f_n).
\]
In particular, this recovers the ring structure of $C^\infty(M)$ upon taking the map $g : \R^2 \to \R$ to be given by the addition or multiplication of two real numbers. Now if we consider the elements of $C^\infty(\R^n)$ as the $n$-ary operations in this way, then plugging these functions into each other results in an infinite system of equations satisfied by these operations. This results in an algebraic theory of which we can then consider arbitrary models: these are sets coming with $n$-ary operations indexed by all smooth functions $\R^n \to \R$ and satisfying the relevant equations; these algebraic structures are the \emph{$C^\infty$-rings}~\citep{MR}. As their name suggests, every $C^\infty$-ring is in particular a ring, obtained by forgetting all algebraic structures apart from addition and multiplication; and for every manifold $M$, the smooth real-valued functions on $M$ form a $C^\infty$-ring $C^\infty(M)$ in a coordinate-independent way.

We thus have two distinct but closely related answers to \ref{which_algebraic}: one which is very much in the spirit of abstract algebra and algebraic geometry by involving only finitely many operations and equations between these; and a more principled one closer to analysis, involving infinitely many defining operations. For the sake of physics, there is no need to prematurely settle on either of these options. Although we will focus on technically developing the second type of approach only, we would prefer to maintain some flexibility and keep in mind that there may yet be further options with other nuances.\footnote{It seems plausible that such open-ended methodology can help explain the low standards of mathematical precision commonly encountered in physics, but we will not dwell on this point.}

We now consider the question \ref{which_physics}. For concreteness, let us assume that we want to formulate general relativity algebraically.
Central to general relativity are Einstein's field equations, which govern how the curvature of spacetime is determined by the distribution of stress-energy density. 
These are standardly expressed in the following form, with coordinate indices $\mu,\nu = 0,\ldots,3$ for the temporal and spatial dimensions:\footnote{For convenience, we use units in which the speed of light is $1$ and the gravitational constant is $1/8\pi$, as is common practice. We also pretend that the cosmological constant is zero for simplicity, but this will not affect our discussion.}\textsuperscript{,}\footnote{We refer to \citet{BD} for an intuitive explanation of the field equations and their phenomenological significance.}
\beq
	\label{efe}
	R_{\mu\nu}-\frac{1}{2} R g_{\mu\nu} =T_{\mu\nu}.
\eeq

In this expression, $R_{\mu\nu}$ is the Ricci tensor, which measures the dynamically relevant spacetime curvature generated by the metric tensor $g_{\mu\nu}$.

$R$ is the Ricci scalar---the contraction of the Ricci tensor to a scalar---and $T_{\mu\nu}$ is the stress-energy tensor that encodes the distribution of matter in spacetime.

To formulate the field equations, Geroch considered algebraic definitions of the technical notions appearing in the field equation~\Cref{efe}, including tensor calculus, metric tensors, the Levi-Civita connection and the Ricci tensor. The relevant structure of $C^\infty(M)$ can be considered to be the minimal one which lets us define these notions. As presented in detail in~\citet{pessers}, they can be defined for any commutative ring satisfying a suitable regularity condition (that its module of derivations has a dual together with which it forms a dual pair). Hence we can interpret the field equations~\Cref{efe} in any such ring, and following Geroch's convention, we can call an \emph{Einstein algebra} any such ring together with a solution to the field equations.

Note that the fact that the algebraic approach starts from manifolds and is adequate for doing physics does not mean that it is necessarily equivalent to the manifold approach. For example, \citet{MR} proposed models for synthetic calculus\footnote{This synthetic calculus is called \emph{smooth infinitesimal analysis}. It is a foundation for calculus alternative to standard analysis and has infinitesimals that square to zero~\citep{MR,bell}.} which are built on the category of $C^\infty$-rings of a certain sort. Objects in this category include quotient rings of $C^\infty(M)$ (with respect to ideals of a certain sort), which includes the ring of dual numbers $C^\infty(\R)/\langle x^2\rangle$.  We can interpret this category geometrically by considering its dual category, which is like the category of smooth manifolds but has extra objects. For example, $C^\infty(\R)/\langle x^2\rangle$ does not correspond to a smooth manifold but to an infinitesimal object in the dual category. This infinitesimal object behaves like a tangent space, but it can be embedded into other objects that are manifolds and therefore can be seen as an infinitesimal part of space (see \citet{sdg_paper} for more details).\footnote{Note that if we wanted to exclude rings like $C^\infty(\R) / \langle x^2 \rangle$ from the category of $C^\infty$-rings, so that the dual category does not have infinitesimal objects, then we would need to impose unnatural restrictions on $C^\infty$-rings. See \citet{RBW} for such a restriction and \citet{sdg_paper} for further comments.} But according to the manifold approach, a manifold does not have infinitesimal parts. In this sense, the algebraic approach is not equivalent but more expressive than the manifold approach.

We now pose the conceptual problems which will motivate the developments proposed in the upcoming sections. Granting that the formalism of Einstein algebras is technically adequate for doing physics up to general relativity, an immediate question is how we should interpret the fundamental algebraic structure of a commutative ring or $C^\infty$-ring. The elements of $C^\infty(M)$ are typically interpreted as the possible field configurations of a scalar field. However, such an algebraic ontology is unmotivated: if the scalar fields are actual physical fields, then it is arbitrary for one such scalar field to be ontologically privileged, as opposed  to other non-scalar physical fields like the electromagnetic field. On the other hand, if the elements of the ring under consideration do not play the role of configurations of any actual physical fields, then they would be strange platonic structures at the fundamental level. We now turn to the question of how to improve on this situation.

Finally, let us comment on the relation between questions \ref{which_algebraic} and \ref{which_physics}. Our main interest is in answering question \ref{which_physics}, but answering \ref{which_algebraic} is an important step for making progress on \ref{which_physics}. For how else would we have come up with the idea of using a structure like commutative rings in the first place? Considering the second question separately also helps with weeding out the redundant structures. As it will become clear later, many derived notions involved in Geroch's approach are not necessary.

\section{Spacetime algebraicism based on physical fields}

Having arrived at the conclusion that spacetime algebraicism based on scalar-valued functions is problematic, we now discuss how one can develop versions of algebraicism in which all physical fields feature among the primitive structures. This constitutes both a conceptual and a technical development, with the latter leveraging definitions and constructions which are known in the mathematical literature.

Thus consider a physical field, say the electromagnetic field, or more generally a collection of physical fields, say the electromagnetic field together with a spinor field. (We will speak in the singular in the following, and leave the straightforward extension to the case of multiple fields largely to the examples.) Given this field, we consider the following questions as constituting a good methodology for determining the algebraic approach to the physics of the field without reference to spacetime points. They extend and generalize the two questions considered in the previous section.
\begin{enumerate}[label=(\arabic*')]
	\setcounter{enumi}{-1}
	\item\label{which_structure2} What manifold-theoretic structures are relevant for formulating the physical field?
	\item\label{which_algebraic2} Which algebraic structures can be defined on the field so formulated?
	\item\label{which_physics2} Which of these algebraic structures are relevant for formulating the physics of the field, and what does a purely algebraic formulation of the physics look like?
\end{enumerate}
Again, \ref{which_physics2} is our main interest and where we move away from the manifold context, while we consider \ref{which_structure2} and \ref{which_algebraic2} as important steps towards answering \ref{which_physics2}. Note that in the case of a single scalar field, the answer to \ref{which_structure2} is clear: the relevant manifold-theoretic structure is the scalar-valued functions over the manifold. This is why we did not consider this question in the previous section. It also shows that in the scalar field case, question \ref{which_algebraic2} reproduces question \ref{which_algebraic} from the previous section, and similarly \ref{which_physics2} reproduces \ref{which_physics}. Hence the above questions directly generalize the ones we had addressed in the scalar field case. Note that sorting out the manifold-theoretic structures in answering question \ref{which_structure2} does not commit us to manifolds ontologically. It is only a methodological starting point. The manifold-theoretic structures will be reconceptualized by algebraic ones in answering question \ref{which_physics2}.

We will examine these general questions in turn in the following sections and discuss methods for how to answer them. These will be illustrated with a few running examples including the electromagnetic field and a Weyl spinor field.

\section{A general manifold-theoretic formulation of physical fields}
\label{geom_structures}

In the case of a scalar field, it was clear that its algebraic description would have to be based on the scalar-valued functions. For other types of fields, it is far less clear what the most appropriate mathematical structure is in response to question \ref{which_structure2}. For example for the electromagnetic field, there are many different kinds of manifold-theoretic structures which have been entertained to describe it:
\begin{itemize}
	\item In the non-relativistic context, the electric and magnetic field are described separately, each formalized by a vector field on space.\footnote{More precisely, with the electric field a vector field and the magnetic field a \emph{pseudovector} field.}
	\item In the relativistic setting, these two vector fields become unified to the antisymmetric \emph{field strength tensor} $F_{\mu\nu}$, or equivalently to a \emph{differential $2$-form} $F$, on spacetime.
	\item Taking the gauge theory perspective, the electromagnetic field is described by the \emph{four-potential} $A_\mu$, or equivalently by a \emph{differential $1$-form} $A$, modulo gauge transformations.\footnote{Technically this $1$-form takes values in the Lie algebra $\mathfrak{u}(1)$, but since this Lie algebra is canonically isomorphic to $\R$, we consider it as an ordinary $1$-form for simplicity.}
	\item In more abstract geometric terms, a field configuration is given by a connection on a \emph{principal bundle} with gauge group $U(1)$.\footnote{Note that on topologically nontrivial spacetimes, these formalisms are not equivalent, with the final one usually considered to be more physically correct; see e.g.~\citep[Section~10.5]{nakahara}.}
\end{itemize}
The technical details in these descriptions are secondary to our point, which is to illustrate the apparent malleability of the identification between physical fields and various manifold-theoretic structures. The situation with general relativity is even worse than this: there are a plethora of reformulations, many of which do not even involve a metric tensor~\citep{krasnov}. 

So how do we answer the question about the proper manifold-theoretic structure describing a physical field then? Should any of the above prescriptions be given priority over the others? Our inclination is to pursue a more general manifold-theoretic formulation of fields, one which does not require one to choose between the various descriptions presented above. (See \Cref{pure_electro1} for more details.)

To motivate the upcoming definition, it is useful to recall an important principle often invoked by physicists:
\begin{center}\smallskip
	\emph{A field is characterized by how it transforms under changes of coordinates.}\smallskip
\end{center}
In order to avoid unnecessary talk about coordinates, we find it advantageous to move from considering the transformation behavior under passive coordinate changes to considering the transformation behavior under active diffeomorphisms. And in fact, for many types of fields it is possible and useful to consider their transformation behavior under more general maps, such as \emph{local diffeomorphisms}, which include (among other things) the coordinate charts of any manifold $M$, in the form of inclusion maps of open subsets of $\R^d$ into $M$.

We prescribe the following:
\begin{itemize}
	\item There is a class of manifolds\footnote{As will become clear in the upcoming discussion, we will focus on manifolds of a fixed spacetime dimension $d$, which will turn out to be necessary for our treatment of Lagrangian densities.}
		\!\!\textsuperscript{,}\!\!
		\footnote{Note that many types of fields considered in physics are not defined on plain manifolds, but rather on manifolds equipped with a certain \emph{background field}, for example a metric. In cases like this, the class of manifolds needs to be chosen so as to take this into account, and consist for example of the \emph{Lorentzian} manifolds. Similarly, the class of designated maps in the second item must be defined so as to consist of those maps which preserve the background fields. (However, see \Cref{lagrangian_preserving} for a counterpoint.)} (or other structures playing an analogous role, such as formal duals of commutative algebras), such that for every manifold $M$ in this class we have a set of field configurations $\ff(M)$.
	\item There is a class of designated maps between these manifolds, playing the role of active transformations, such that for every map $f : M \to N$ in this class, we have a designated function $\ff(f) : \ff(N) \to \ff(M)$, which specifies the transformation behavior of the field by mapping field configurations on $N$ to field configurations on $M$. This class is closed under composition and containing all identity maps.
		
\end{itemize}
The consistency of field transformations requires that these prescriptions make $\ff$ into a \emph{functor}: it must be compatible with composing active transformations by satisfying the equations
		\beq
			\label{f_functoriality}
			\ff(g \circ f) = \ff(f) \circ \ff(g), \qquad \ff(\id) = \id.
		\eeq
Since every diffeomorphism can be used in place of $f$, the functor $\ff$ in particular specifies how the field transforms under diffeomorphisms.

Our discussion thus far has been concerned with the \emph{kinematical} aspects of a field by focusing on the possible field configurations and how these transform. But certainly the specification of a physical field must also involve some \emph{dynamical} information. The dynamical information about the field is usually taken to come in the form of its \emph{Lagrangian density},\footnote{This is the case at least for \emph{local} interactions, which we restrict ourselves to in this paper.} both in classical as well as in quantum field theory. But before we formalize this, we need to consider what a \emph{density} is in general. A density $\omega$ on a $d$-dimensional manifold $M$ assigns to every family of tangent vectors $v_1,\ldots,v_d \in T_x M$ at any point $x \in M$ a number $\omega(v_1,\ldots,v_d)$, to be interpreted as the volume of the infinitesimal parallelepiped spanned by $v_1, \ldots, v_d$, such that suitable linearity and smoothness conditions hold which formalize this intended interpretation~\citep{nicolaescu}. Let us write $\df(M)$ for the set of all densities on the manifold $M$. If $M$ and $N$ are manifolds of the same dimension $d$ and $f : M \to N$ is any smooth map, then a density on $N$ can be pulled back along $f$ to a density on $M$, by declaring the volume of an infinitesimal parallelepiped to be the volume of its image in $N$. In this way, we obtain a map $\df(f) : \df(N) \to \df(M)$, making the functoriality equations \eqref{f_functoriality} hold for $\df$ as well. In other words, between manifolds of the same dimension, densities display the same transformation behavior as a physical field.

A Lagrangian density $\LD$ now assigns to every field configuration $\phi \in \ff(M)$ on every manifold $M$ a density $\LD(\phi) \in \df(M)$. In other words, the dynamical content of the field is encoded in a \emph{transformation} $\LD : \ff \to \df$, by which we mean a family of maps $(\LD_M)$ having components of the form
\[
	\LD_M : \ff(M) \to \df(M)
\]
for all manifolds $M$ of fixed dimension $d$. The standard postulate that the Lagrangian density should be coordinate-independent now translates into the requirement that for every map $f : M \to N$ in the designated class, the diagram
\beq
	\label{LDnatural}
	\begin{tikzcd}
		\ff(N)	\ar{r}{\LD_N} \ar{d}[swap]{\ff(f)}	& \df(N)	\ar{d}{\df(f)}	\\
		\ff(M)	\ar{r}{\LD_M}				& \df(M)
	\end{tikzcd}
\eeq
commutes: applying the transformation first and then evaluating the Lagrangian density must be the same as evaluating the Lagrangian density first and then transforming the resulting density. This amounts to the requirement that the map $f$ is not only a map with respect to which the field under consideration has definite transformation behavior, but that it actually is a \emph{physical symmetry}\footnote{Note that a ``symmetry'' in this sense does not need to be invertible (\Cref{lagrangian_preserving}).} preserving the Lagrangian.

It is worth emphasizing that the current references to manifolds is a methodological step towards formulating algebraic structures of fields. In our proposal in \Cref{sec_natural,sec_physics}, it will be clear that we do not posit manifolds in our ontology. In Geroch's approach, the reference to manifolds appears in exactly the same form: it motivates the consideration of commutative rings in general by noting that $C^\infty(M)$ is a commutative ring for every manifold $M$.

By putting these pieces together, we propose the following definition of a field.

\begin{defn}
	\label{field_defn}
	Let $\ST$ be a category of (conceivable) spacetimes of fixed dimension $d$, with maps with respect to which the field in question has definite transformation behavior. Then we identify a field with a pair $(\ff,\LD)$ consisting of the following:
	\begin{enumerate}
		\item A \emph{field functor}\footnote{The superscript ``$\mathsf{op}$'' here indicates that the functor is \emph{contravariant}, by which one means that it reverses the order of composition of maps, as indicated in the functoriality equation~\eqref{f_functoriality}.}
			\[
				\ff : \ST^\op \to \Set,
			\]
			assigning to every spacetime the set of field configurations on it, and to every map of spacetimes the field's transformation behavior.
		\item A natural transformation $\LD : \ff \to \df$, meaning a map
			\[
				\LD_M : \ff(M) \to \df(M)
			\]
			for every $M \in \ST$ assigning to every field configuration its Lagrangian density, such that the diagram~\eqref{LDnatural} commutes for every map $f$ in $\ST$.
	\end{enumerate}
\end{defn}

As we will illustrate in \Cref{spinor_electro1}, this definition can also deal with two or more physical fields at the same time, by considering them jointly as a single field.

Note that this definition is schematic, in the sense that we have not specified the technical details of how the category $\ST$ can or should be defined. A more precise prescription to arrive at a canonical choice is discussed as follows.

\begin{rem}
	\label{lagrangian_preserving}
	As a first example, consider the case where the category $\ST$ only contains Minkowski spacetime as its single object and (active) Lorentz transformations as the morphisms. Then, a physical field according to our definition must have definite transformation behavior with respect to Lorentz transformations, and its Lagrangian density must be Lorentz invariant. \Cref{field_defn} generalizes this to the case where the field may be defined on an entire class of spacetimes, and the allowed transformations no longer need to be invertible, although one may still think of them as physical symmetries. In this way, our definition is more general and expressive than ordinary Lorentz invariance.

	For a particular physical field such as the electromagnetic field, there may be some ambiguity in how to construct the relevant category $\ST$. But there is a canonical choice given as follows.
	The collection of all $d$-dimensional manifolds, equipped with all the background fields needed to define the Lagrangian density, forms a category together with \emph{all} the maps with respect to which the given field has any definite transformation behavior at all. Let us denote it $\ST_{\max}$, to indicate that the class of maps is now (informally) the maximal category on which the field functor can be defined. 

	Then in general, the naturality square~\eqref{LDnatural} does not commute, since having definite transformation behavior under a map $f$ does not imply that this transformation preserves the field's Lagrangian density. A physical symmetry $f$, or we could call it a \emph{Lagrangian-preserving map} $f$, is then defined to be a map in $\ST_{\max}$ for which the naturality square~\eqref{LDnatural} \emph{does} commute. It is easy to see that the Lagrangian-preserving maps are also closed under composition and contain all identities. We thus propose to define $\ST$ as the subcategory of Lagrangian-preserving maps. The commutativity of \eqref{LDnatural} on $\ST$ then holds by definition.
	One of the consequences of this proposal is that a field theory is generally covariant if and only if every map in $\ST_{\max}$ is already Lagrangian-preserving, so that $\ST = \ST_{\max}$.

	As we will see in the following examples, the Lagrangian-preserving maps are typically, but not always, those which preserve all the ``background fields'' which appear in the Lagrangian, and this concept generalizes the usual notion of physical symmetries to not necessarily invertible maps.
\end{rem}

\begin{scalar_ex}
	\label{pure_scalar1}
	For a real-valued scalar field, the set of field configurations on a manifold $M$ is precisely the set of real-valued smooth functions on $M$, so that $\ff(M) = C^\infty(M)$. This has in principle well-defined transformation behavior with respect to \emph{any} smooth map between manifolds $f : M \to N$. Indeed a scalar field configuration $\phi \in C^\infty(N)$ can be pulled back along a smooth map $f : M \to N$, simply by composing $f$ with $\phi : N \to \R$ to $\phi \circ f : M \to \R$, regardless of whether $M$ and $N$ have the same dimension. So at the purely kinematical level, we could work with the category consisting of all smooth manifolds and smooth maps between them. On this category, our field functor is $\ff = C^\infty$, for which $C^\infty(f) : C^\infty(N) \to C^\infty(M)$ is given by composition with $f$.

	To take the dynamics into account, let us consider massless\footnote{All of our considerations apply likewise in the presence of a mass term, so the masslessness assumption merely serves to simplify the appearance of our equations.} $\phi^4$ theory as a basic example. Its Lagrangian density transformation is given by $\LD^{\mathrm{scalar}}$ with components
	\beq
		\label{Lscalar}
		\LD^{\mathrm{scalar}}_M(\phi) = \left( \frac{1}{2} g^{\mu \nu} \, (\partial_\mu \phi) \, (\partial_\nu \phi) - \frac{\lambda}{4!} \phi^4 \right) \sqrt{|\det g|},
	\eeq
	where $(g^{\mu\nu})$ is the metric tensor on the spacetime manifold $M$---typically assumed to be of Lorentzian signature---the coupling constant $\lambda$ is an arbitrary (but fixed) parameter of the theory, and $\sqrt{|\det g|}$ is the usual density associated with the metric $g$.

	So then let us determine the Lagrangian-preserving maps. To begin with, the objects of $\ST_{\max}$ need to be $d$-dimensional \emph{Lorentzian} manifolds which are each equipped with a metric tensor $(g^{\mu\nu})$ of Lorentzian signature (which, as we will explain in \cref{gr1}, could also be construed as a physical field in our setting). Since the transformation behavior of the field is independent of the Lagrangian density, we can still consider the maps in $\ST_{\max}$ to be all smooth maps, regardless of whether they are compatible with the given metrics.

	Now if $(M,g)$ and $(N,h)$ are arbitrary Lorentzian manifolds and $f : M \to N$ is an arbitrary smooth map, then the commutativity of the naturality diagram~\eqref{LDnatural} becomes the requirement that
	\begin{align*}
		\bigg( \frac{1}{2} g^{\alpha\beta}	\left[ \frac{\partial \phi}{\partial y^\mu}(f(x)) \frac{\partial f^\mu}{\partial x^\alpha}(x) \right]	&
						\left[ \frac{\partial \phi}{\partial y^\nu}(f(x)) \frac{\partial f^\mu}{\partial x^\beta} (x) \right]	
						- \frac{\lambda}{4!} \phi^4(f(x))	\bigg) \sqrt{|\det g(x)|}	\\[4pt]
		& = \bigg( \frac{1}{2} h^{\mu\nu} \left[ \frac{\partial \phi}{\partial y^\mu} (f(x)) \right] \left[ \frac{\partial \phi}{\partial y^\nu}(f(x)) \right] - \frac{\lambda}{4!} \phi^4(f(x)) \bigg) \sqrt{|\det h(f(x))|}
	\end{align*}
	for all $\phi \in C^\infty(N)$ and all $x \in M$. One can check that $f$ is therefore Lagrangian-preserving if and only if it preserves the metric, in the sense that $g = f^*(h)$ with metrics pulling back along $f$ as usual (and as explained in the upcoming \Cref{gr1}). This implies in particular that $f$ must be a local diffeomorphism. So for example when both $(M,g)$ and $(N,h)$ are given by Minkowski spacetime, then the Lagrangian-preserving maps are precisely the Lorentz transformations, since these are well-known to be precisely the special-relativistic physical symmetries of massless $\phi^4$-theory. 

	So to summarize, for massless $\phi^4$ theory in $d$ dimensions, a reasonable choice is to take the category $\ST$ to have $d$-dimensional Lorentzian manifolds as objects and metric-preserving local diffeomorphisms as morphisms, the field functor is $\ff = C^\infty$, and the Lagrangian density transformation is given by~\eqref{Lscalar}, which indeed maps every $\phi \in C^\infty(M)$ to some density $\LD(\phi) \in \df(M)$ for every Lorentzian manifold $(M,g)$ and the naturality condition holds by construction.
\end{scalar_ex}

\begin{electro_ex}
	\label{pure_electro1}
	Consider pure electrodynamics as the second example. Then one can use either of the four standard formulations of the electromagnetic field outlined at the beginning of this section in order to define the field configurations on any manifold $M$, and thereby the field functor $\ff$. As long as one includes only contractible manifolds in $\ST$, then these formulations result in isomorphic field functors, which shows that these formulations are equivalent (at least on topologically trivial manifolds). It is in this sense that the notion of field functor is more general than those standard formulations, and does not require us to choose between them.
	
	For the sake of concreteness, we take the field configurations to correspond to the differential $1$-forms, and suppose that we ignore gauge equivalence for the moment. Then we have
	\[
		\ff(M) \, = \, \Omega^1(M).
	\]
	If $M$ and $N$ are manifolds and $f : M \to N$ is any smooth map, then any $1$-form $A$ on $N$, representing a configuration of the electromagnetic field on $N$, can be pulled back along $f$ to a $1$-form $f^*(A)$ on $M$, representing a configuration of the electromagnetic field on $M$. This specifies the transformation behavior of the electromagnetic potential, or in other words the action of the field functor $\ff$ on morphisms.
	
	In pure electrodynamics, the usual Lagrangian density on a spacetime manifold $(M,g)$ is given by
	\beq
		\label{Lem}
		\LD^{\mathrm{em}}_M(A) = - \frac{1}{4} g^{\alpha\mu} g^{\beta\nu} F_{\alpha\beta} F_{\mu\nu} \sqrt{|\det g|}	\qquad\quad (\textrm{with } F_{\mu\nu} = \partial_\mu A_\nu - \partial_\nu A_\mu),
	\eeq
	which is more commonly written as $-\frac{1}{4} F_{\alpha\beta} F^{\alpha\beta} \sqrt{|\det g|}$ or in terms of differential form notation and the Hodge star operator simply as\footnote{However, it should be noted that the expression $-\frac{1}{4} F \wedge \star F$ is not entirely equivalent to \eqref{Lem}, since it requires an identification of densities with volume forms via an orientation, which is assumed in most standard treatments of Lagrangian field theory but requires one to be working with oriented Lorentzian manifolds instead of mere Lorentzian manifolds.} $-\frac{1}{4} F \wedge \star F$. In our \eqref{Lem}, we have written out the raising of the indices via the metric in order to make the dependence on the metric completely explicit.

	It follows that, just as in \Cref{pure_scalar1}, the objects of our category should again be $d$-dimensional Lorentzian manifolds, and we can consider $\ST_{\max}$ to contain all smooth maps between these, since differential $1$-forms then have definite transformation behavior.
	The Lagrangian-preserving maps can now be determined either by a messy coordinate calculation, or alternatively by a more abstract argument along the following lines. Up to the constant factor of $-\frac{1}{4}$, the Lagrangian density~\eqref{Lem} is given by the scalar-function-valued inner product of the field strength tensor $F = dA$ with itself times $\sqrt{|\det g|}$, where the inner product on $2$-forms is the one induced from the inner product of vectors. Since the exterior derivative commutes with pullback $f^*$ of differential forms, it follows that a smooth map $f : M \to N$ between Lorentzian manifolds $(M,g)$ and $(N,h)$ is Lagrangian-preserving if and only if the equation
	\[
		\langle f^*(dA), f^*(dA) \rangle_M(x) \, \sqrt{|\det g(x)|} \, = \, \langle dA, dA \rangle_N(x) \, \sqrt{|\det h(f(x))|}
	\]
	holds for all $A \in \Omega^1(N)$ and $x \in M$. In $d = 4$ spacetime dimensions, it can then be shown that $f$ is Lagrangian-preserving if and only if it is \emph{conformal}, which means that there exists a scalar function $c \in C^\infty(M)$ such that $f^*(h) = c \hspace{1pt} g$. This recovers the well-known conformal invariance of electrodynamics~\citep{CO}.

	Thus one sensible choice for the category $\ST$ would have four-dimensional Lorentzian manifolds as objects and \emph{conformal} maps between these as morphisms, where the field functor is the functor of differential $1$-forms $\Omega^1$.
	The Lagrangian density \eqref{Lem} specifies a natural transformation $\LD^{\mathrm{em}} : \Omega^1 \to \df$, and therefore defines the electromagnetic field in the setting of \Cref{field_defn}, so far without gauge equivalence.
	
	To take gauge equivalence into account, we need to consider equivalent any two field configurations $A \in \Omega^1(M)$ and $A' \in \Omega^1(M)$ for which there is a scalar function $\lambda \in \Omega^0(M)$ such that
	\[
		A' = A + d\lambda.
	\]
	In other words, the set of \emph{physical} field configurations is then given by the quotient vector space
	\[
		\ff(M) \, = \, \Omega^1(M) / d \Omega^0(M),
	\]
	where $d \Omega^0(M)$ denotes the vector space of all $1$-forms which are differentials of scalar functions on $M$.
	The transformation behavior $\ff(f) : \ff(N) \to \ff(M)$ is still the same as above, meaning that an equivalence class $[A] \in \ff(N)$ represented by some $A \in \Omega^1(N)$ gets mapped to the equivalence class $[f^*(A)] \in \ff(M)$, which is well-defined for any smooth map $f$ since the exterior derivative $d$ and the pullback of differential forms $f^*$ commute, $f^*(d\lambda) = df^*(\lambda)$. Since the Lagrangian density~\eqref{Lem} is gauge invariant, it still defines a transformation $\LD : \ff \to \df$ which is natural precisely with respect to the conformal maps (assuming four spacetime dimensions).

	Note that taking gauge invariance into account only modifies the field functor, while $\ST$ and the Lagrangian density transformation remain unchanged.
\end{electro_ex}

\begin{gr_ex}
	\label{gr1}
	For general relativity in $d$ spacetime dimensions, one may again consider all $d$-dimensional manifolds as the a priori conceivable spacetimes making up the objects of the category $\ST$. Now we want to consider a metric of Lorentzian signature together with the Einstein-Hilbert Lagrangian as our physical field. Therefore $\ff(M)$ should be the set of all Lorentzian metrics on any given $d$-dimensional manifold $M$.\footnote{Note that there are manifolds $M$ which do not admit any Lorentzian metric, in which case $\ff(M)$ is the empty set.}  The metric tensor can be pulled back under local diffeomorphisms, but not under general smooth maps (due to the non-degeneracy requirement\footnote{Alternatively, one could consider a modification of general relativity which allows for the metric to be degenerate, and perhaps even of arbitrary signature. Indeed it has been argued that allowing degeneracies is even preferable over the standard formulation by allowing the possibility to include singularities in spacetime \citep{stoica}. In this case, the metric would have well-defined transformation behavior with respect to all smooth maps, which is what had happened also in the previous two examples.}). It is thus natural to include only local diffeomorphisms as the morphisms in $\ST$.
	Like this, we obtain a field functor $\ff : \ST^\op \to \Set$ in a similar way as before. The Lagrangian density is the one corresponding to the Einstein-Hilbert action as usual,
	\[
		\LD^{\mathrm{GR}}_M = (R - 2 \Lambda) \sqrt{|\det g|},
	\]
	where $R$ is the Ricci curvature scalar (depending on $g$) and $\Lambda$ the cosmological constant.

	Note that in this case, the distinction between general maps with respect to which the field transforms and the more specific Lagrangian-preserving maps is not needed, since the Lagrangian preservation is automatic. This property captures the general covariance of general relativity.
\end{gr_ex}

\begin{spinor_ex}
	\label{pure_spinor1}
	Let us consider the field theory of a (necessarily massless) Weyl spinor in four spacetime dimensions. This means that the conceivable spacetimes are the four-dimensional Lorentzian manifolds $(M,g)$, where now the metric already enters into the definition of the set of field configurations, through the construction of the Clifford bundle associated to $M$ via the metric $g$ and its associated spinor bundles~\citep{LM}. Strictly speaking, a spinor field configuration consists of a spin structure \emph{plus} a section of the corresponding spinor bundle.

	In this case, finding the maximal class of maps with respect to which the field has definite transformation behavior is a surprisingly subtle problem, where contradictory statements have coexisted in the literature for decades~\citep{pitts}. Hence the general methodology of restricting to the Lagrangian-preserving maps (\Cref{lagrangian_preserving}) encounters limitations. Nevertheless, it is clear that defining $\ST$ to consist of Lorentzian manifolds together with metric-preserving maps is a reasonable choice, and makes the Weyl Lagrangian density
	\[
		\LD^{\mathrm{Weyl}}_M(\psi) := i \overline{\psi} D \psi,
	\]
	where $D := \sigma^\mu \partial_\mu$ is the Dirac operator on the spinor bundle on which $\psi$ is defined, into a natural transformation. Note that we suppress the spin structure as part of the specification of the field configuration, although it technically also needs to be treated as part of it.
\end{spinor_ex}

When considering several physical fields at the same time, we can either deal with a corresponding number of field functors at once, or consider them jointly as a single field functor as follows.

\begin{spinor_electro_ex}
	\label{spinor_electro1}
	Combining the electrodynamics and Weyl spinor examples, let us consider \emph{Weyl spinor electrodynamics}, which contains both of those fields coupled together. We thus have two field functors, where $\ST$ again consists of Lorentzian manifolds and smooth maps between them, given by $\ff_1 = \Omega^1$ and $\ff_2$ the Weyl spinor field functor, with their separate Lagrangian densities. We now explain how these two fields can be jointly regarded as a \emph{single} field, in the sense of \Cref{field_defn}, while taking gauge equivalence into account.

	To begin, recall that a gauge transformation for these fields on a spacetime manifold $M$ is determined by some real-valued scalar function $\lambda \in C^\infty(M)$. Denoting the electromagnetic field by $A$ and the spinor field by $\psi$, the transformation is
	\[
		A' := A + d\lambda, \qquad \psi' := e^{i\lambda} \psi.
	\]
	This suggests that these two fields should actually be considered as separate components or aspects of a single field, with sets of field configurations given by
	\beq
		\label{gauge_joint}
		\ff(M) \, := \, \left[ \ff_1(M) \times \ff_2(M) \right] \,/\, \left[ \textrm{modulo gauge equivalence} \right].
	\eeq
	This definition indeed has well-defined transformation behavior under every map $f : M \to N$ in $\ST$, since this holds for both components fields and these transformations respect gauge equivalence. Therefore $\ff$ is indeed a field functor describing the kinematics of both fields jointly in a unified way. The Lagrangian $\LD^{\mathrm{tot}} : \ff \to \df$ is then the sum of the individual Lagrangians, where the Dirac operator acting on the Weyl spinor needs to be gauged as usual in order to achieve gauge invariance. Formally, on every Lorentzian manifold $(M,g)$ we have
	\[
		\LD^{\mathrm{tot}}_M(A,\psi) := \left( - \frac{1}{4} g^{\alpha\mu} g^{\beta\nu} F_{\alpha\beta} F_{\mu\nu} + i \overline{\psi} D_{\mathrm{\scriptscriptstyle{gauged}}} \psi \right) \sqrt{|\det g|},
	\]
	where the Dirac operator now takes the form
	\[
		D_{\mathrm{\scriptscriptstyle{gauged}}} := \sigma^\mu \left( \partial_\mu - i A_\mu \right),
	\]
	and the gauge invariance is a standard computation. Due to this gauge invariance, we indeed obtain a map $\LD^{\mathrm{tot}}_M : \ff(M) \to \df(M)$ as desired, and the naturality~\eqref{LDnatural} holds for all metric-preserving maps.
	% Unfortunately, this type of description does not seem to interact well with the constructions of the upcoming sections, and we thus leave the treatment of gauge equivalence in scalar electrodynamics to future work.
\end{spinor_electro_ex}

Finally, we shall remark that our definition of fields does not  axiomatize all properties which are typically satisfied by physical fields (and doing so is not relevant for our purposes). In particular, it does not account for the strong kind of \emph{locality} often displayed by those field functors which describe physical fields, where a field configuration can be ``pieced together'' from a family of configurations on local patches. We refer to the notion of \emph{bundle functor} for a stronger definition which incorporates such locality properties more stringently~\citep[Definition~14.1]{KMS}, and note that imposing suitable \emph{sheaf conditions} on the field functor would have a similar effect, although care must be taken with the treatment of gauge invariance.

\section{Algebraic structure from natural operations}
\label{sec_natural}

We now turn to question \ref{which_algebraic2}, which algebraic structures can be defined on a physical field formulated manifold-theoretically? In light of the previous section, the question becomes this: which algebraic structure does a field in the sense of \Cref{field_defn} carry? More precisely: what is the richest algebraic structure that can be defined on the field configurations in a way which respects the symmetries?
The following technical developments will address this question, considering algebraic operations of finite arity.\footnote{A technically distinct but closely related answer could be given based on the \emph{codensity monads} of category theory~\citep{leinster}, in which operations of infinite arity are allowed. See \href{https://golem.ph.utexas.edu/category/2020/01/codensity_monads.html\#c057683}{https://golem.ph.utexas.edu/category/2020/01/codensity\_monads.html\#c057683} for some discussion on the scalar field case.}
Our goal in \Cref{sec_physics} will then be to reaxiomatize a field (or multiple fields) as an algebraic structure of that type, no longer referring to any spacetime manifold.

\subsection{Natural operations}

Algebraic structures are defined in terms of \emph{algebraic operations}, which are ways of combining elements of a set such as to obtain another element.  In our context, one can add or multiply two scalar-valued functions on a manifold so as to obtain a third; one can add two differential $1$-forms so as to obtain a third; or in a multi-sorted setting where both of these structures appear, one can take the derivative of a scalar-valued function so as to obtain a differential $1$-form.

The characteristic feature of these type of operations is that they are coordinate-independent.\footnote{Note that a coordinate-independent construction may still invoke coordinates, as long as its result is independent of the particular choice.} In the previous section, we have generalized the notion of coordinate independence to that of definite transformation behavior under a specified class of maps, which typically includes local diffeomorphisms. We consider it more elegant to work in this setting.  For example, for the addition of scalar-valued functions, the fact that it has the relevant invariance properties then amounts to the fact that for every smooth map $f : M \to N$ between manifolds, the diagram
\[
	\begin{tikzcd}
		C^\infty(N) \times C^\infty(N) \ar{r}{+} \ar[swap]{d}{C^\infty(f) \times C^\infty(f)}		& C^\infty(N) \ar{d}{C^\infty(f)}	\\
		C^\infty(M) \times C^\infty(M) \ar{r}{+}							& C^\infty(M)
	\end{tikzcd}
\]
commutes: if we add to smooth functions on $N$ and then pull the result back to $M$, that gives the same function as first pulling them back individually. It is clear that this specializes to ordinary diffeomorphism covariance in the case where $f : M \to N$ is a diffeomorphism.

The general theory behind this type of operation is that of \emph{natural operations} of differential geometry~\citep{KMS}.
We next state a simplified variant of the definition used in~\citep{KMS}.
For $n \in \N$ and a field functor $\ff : \ST^\op \to \Set$, we can consider the field functor $\ff^{\times n} : \ST^\op \to \Set$ which associates to every manifold $M$ the $n$-fold Cartesian product set
\[
	\ff(M) \times \cdots \times \ff(M),
\]
where the operation of pulling back tuples is defined componentwise. For example, the $n$-fold product of the real scalar field functor $M \mapsto C^\infty(M)$ is isomorphic to the functor which assigns to every manifold $M$ the set of $n$-tuples of smooth functions $M \to \R$, or equivalently the set of smooth functions $M \to \R^n$.

\begin{defn}
	Let $\ff : \ST^\op \to \Set$ be a field functor and $n \in \N$. A \emph{natural operation} of arity $n$ on $\ff$ is a natural transformation $\ff^{\times n} \to \ff$.
\end{defn}

More explicitly, a natural operation $\Phi$ thus consists of maps
\[
	\Phi_M \: : \: \underbrace{\ff(M) \times \cdots \times \ff(M)}_{n \textrm{ times}} \longrightarrow \ff(M)
\]
defined for every manifold $M \in \ST$, such that the diagram
\beq
	\label{naturality}
	\begin{tikzcd}
		\ff(N) \times \cdots \times \ff(N) \ar{r}{\Phi_N} \ar[swap]{d}{\ff(f)^{\times n}}	& \ff(N) \ar{d}{\ff(f)}		\\
		\ff(M) \times \cdots \times \ff(M) \ar{r}{\Phi_M}					& \ff(M)
	\end{tikzcd}
\eeq
commutes for every morphism $f : M \to N$ with respect to which the field has definite transformation behavior.

The natural operations have been classified completely for some manifold-theoretic structures, a fact that one can try to exploit in order to give more concrete descriptions of the algebraic structures that will arise as field algebra. To illustrate the notion mathematically, let us temporarily (only in the following example) relax the requirement in \Cref{field_defn} that the spacetime dimension is fixed and focus just on the kinematic aspect of a scalar field.

\begin{scalar_fun_ex}
	\label{scalar_fun1}

Let $\Mfd$ be the category of manifolds of \emph{all} dimensions, and $C^\infty : \Mfd^\op \to \Set$ the functor which assigns to every manifold the set of real-valued scalar functions on it. This example is illustrative since the natural operations on $C^\infty$ can be completely classified.
	We claim that there is a canonical bijection between the natural operations of arity $n$ and the set $C^\infty(\R^n)$, the smooth functions on Euclidean space $\R^n$. This will be obvious and known to readers who have encountered natural operations before, but we nevertheless present the argument here.

	Indeed in one direction, any $s \in C^\infty(\R^n)$ defines a natural operation given by
	\beq
		\label{Cinfty_natural}
		C^\infty(M) \times \cdots \times C^\infty(M) \longrightarrow C^\infty(M),	\qquad		(\phi_1,\ldots,\phi_n) \longmapsto s \circ (\phi_1,\ldots,\phi_n).
	\eeq
	In words, this simply means that we obtain a new smooth function from given smooth functions $\phi_1,\ldots,\phi_n$ by plugging in their values into the $n$-variable smooth function $s$. It is straightforward to see that this makes the relevant diagram~\Cref{naturality} commute. In the other direction, suppose that we are given a natural operation $\Phi$ for $C^\infty$. Then we can in particular consider its component at the manifold $M = \R^n$ and apply it to the $n$-tuple of functions given by the $n$ coordinate projections $p_1,\ldots,p_n : \R^n \to \R$, which results in the desired smooth function $\Phi_{\R^n}(p_1,\ldots,p_n) \in C^\infty(\R^n)$. It remains to be shown that these two constructions are inverses of each other, for which we sketch the argument in the following paragraph.

	If we start with the smooth function $s$ and construct the associated natural operation, then it is clear that applying it to $p_1,\ldots,p_n$ recovers exactly $s$. Conversely if $\Phi$ is any natural operation, we now show that each one of its components is given exactly by composition with the function $\Phi_{\R^n}(p_1,\ldots,p_n)$. This somewhat surprising fact essentially follows from the \emph{Yoneda lemma} of category theory.\footnote{For any manifold $M$, we can consider given functions $\phi_1,\ldots,\phi_n \in C^\infty(M)$ as the $n$ components of a smooth map $\phi : M \to \R^n$, so that $\phi_i = p_i \circ \phi$. Instantiating the naturality diagram~\Cref{naturality} on $f = \phi$ then gives the commutative diagram
	\[
		\begin{tikzcd}[ampersand replacement=\&]
			C^\infty(\R^n) \times \cdots \times C^\infty(\R^n) \ar{r}{\Phi_{\R^n}} \ar{d}{C^\infty(\phi)^{\times n}}	\& C^\infty(\R^n) \ar{d}{C^\infty(\phi)}		\\
			C^\infty(M) \times \cdots \times C^\infty(M) \ar{r}{\Phi_M}						\& C^\infty(M)
		\end{tikzcd}
	\]
	Starting with the $n$-tuple $(p_1,\ldots,p_n)$ in the upper left gives $(\phi_1,\ldots,\phi_n)$ in the lower left, and then $\Phi_M(\phi_1,\ldots,\phi_n)$ in the lower right. Going the other way results in $\Phi_{\R^n}(p_1,\ldots,p_n)$ in the upper right and then $\Phi_{\R^n}(p_1,\ldots,p_n) \circ \phi$ in the lower right. The commutativity of the diagram thus proves the desired equality
	\[
		\Phi_M(\phi_1,\ldots,\phi_n) = \Phi_{\R^n}(p_1,\ldots,p_n) \circ (\phi_1,\ldots,\phi_n),
	\]
	since $\phi$ is exactly equal to its list of components $(\phi_1,\ldots,\phi_n)$.}
\end{scalar_fun_ex}

\begin{scalar_ex}
	\label{pure_scalar2}
	Consider massless scalar $\phi^4$-theory in $d$ dimensions as introduced before in \Cref{pure_scalar1}. To recall, $\ST$ consists of all $d$-dimensional Lorentzian manifolds, with the designated maps between $(M,g)$ and $(N,h)$ being precisely those smooth maps $f : M \to N$ with $g = f^*(h)$, which are necessarily local diffeomorphisms.

	It is obvious that all the operations of the form~\eqref{Cinfty_natural} are still natural operations in the present setting. 
	But there are also a number of additional natural operations relative to those of \Cref{scalar_fun1}, and we do not know how to determine them all. For example, applying the d'Alembertian to a scalar function gives a natural operation of arity one, 
	\begin{align}
		\begin{split}
			\label{scalar_dalembertian}
			C^\infty	& \longrightarrow	C^\infty							\\
			\phi		& \longmapsto		\Box \hspace{1pt} \phi := g^{\mu\nu} \, \partial_\mu \partial_\nu \phi,
		\end{split}
	\end{align}
	and there is a related natural operation of arity two corresponding to the kinetic term in the Lagrangian,\footnote{Perhaps in contrast to what our notation $\langle d-,d-\rangle$ suggests, this is a primitive operation in this context not built out of separate operations $d$ and $\langle -,- \rangle$.}
	\begin{align}
		\begin{split}
			\label{1form_scalar_product}
			C^\infty \times C^\infty	& \longrightarrow	C^\infty							\\
			(\phi,\psi)			& \longmapsto \langle d \phi, d\psi \rangle := g^{\mu\nu} \, (\partial_\mu \phi) \, (\partial_\nu \psi).
		\end{split}
	\end{align}
	Furthermore, the curvature scalar $R$ defines a natural operation of arity zero, since it produces a coordinate-independent scalar function from no arguments, and likewise for other scalar curvature invariants~\citep{CBCR}.

	Thus there now is a substantial number of natural operations which encode geometric information about the manifolds under consideration. Moreover, another new feature over the case of \Cref{scalar_fun1} is that the set of natural operations now depends on the spacetime dimension $d$, since for example the curvature scalar is distinct from the zero natural operation only for $d \ge 2$. We are not aware of any (even conjectural) description of \emph{all} natural operations in this context, neither for any fixed dimension $d \ge 2$ nor in the ``stable'' range $d \gg 1$~\citep{markl}.
\end{scalar_ex}

\newcommand{\tinyspace}{\hspace{0.7pt}}

\begin{electro_ex}
	\label{pure_electro2}
	Continuing on from \Cref{pure_electro1}, consider first the natural operations on the field functor $\ff = \Omega^1$, corresponding to electrodynamics without taking gauge equivalence into account. Then for any coefficients $c_1,\ldots,c_n$, we obtain a natural operation of arity $n$, given by taking the corresponding linear combination of differential $1$-forms,
	\[
		(\alpha_1,\ldots,\alpha_n) \longmapsto \sum_{i=1}^n c_i \alpha_i.
	\]
	In fact, it is known that these are all the natural operations on $\Omega^1$ when considered as a functor $\Mfd^\op \to \Set$ defined on all manifolds and all smooth maps between them~\citep{NS}, in the style of \Cref{scalar_fun1}.

	Now in $d = 4$ spacetime dimensions, we have seen that the natural choice for $\ST$ contains all the conformal maps as morphisms. With this choice, we have not been able to find any additional natural operations besides the above linear combination operations either.\footnote{As we will see in \Cref{spinor_electro2}, we would obtain a large number of further natural operations if we restricted to metric-preserving maps, and it is possible that some of them can be combined in a way which results in the stronger property of conformal invariance, although we have not found any way of doing so.}

	Upon taking gauge equivalence into account, the field functor becomes $\ff = \Omega^1 / d \Omega^0$. Again, taking linear combinations defines natural operations, and we have not been able to find any other ones.
\end{electro_ex}

\begin{spinor_ex}
	\label{pure_spinor2}
	Consider the theory of a Weyl spinor in four spacetime dimensions from \Cref{pure_spinor1}. It is obvious that taking linear combinations of spinors again results in a family of natural operations of all arities, where now arbitrary complex coefficients are allowed. We also have the application of the Dirac operator, which is a natural operation of arity one,
	\[
		\psi \longmapsto D \psi,
	\]
	as well as a ternary natural operation given by forming the inner product of a spinor with an adjoint spinor, and acting by the resulting scalar on a third spinor,
	\beq
		\label{spinor_ip}
		(\psi, \phi, \chi) \longmapsto (\overline{\psi} \phi) \tinyspace \chi.
	\eeq
	A substantial number of additional operations natural with respect to metric-preserving maps arise and are of a similar flavor. For example, the facts that a bilinear construction like $\overline{\psi} \sigma^\mu \phi$ defines a vector field, the inner product of two vector fields is a scalar function, and a scalar function can act on spinors by multiplication shows that there is a natural operation of arity five given by
	\beq
		\label{fierz1}
		(\psi_1, \phi_1, \psi_2, \phi_2, \chi) \longmapsto (\overline{\psi}_1 \sigma^\mu \phi_1) \tinyspace (\overline{\psi}_2 \sigma_\mu \phi_2) \tinyspace \chi,
	\eeq
	and similarly
	\beq
		\label{fierz2}
		(\psi_1, \phi_1, \psi_2, \phi_2, \chi) \longmapsto (\overline{\psi}_1 \sigma^\mu \sigma^\nu \phi_1) \tinyspace (\overline{\psi}_2 \sigma_\mu \sigma_\nu \phi_2) \tinyspace \chi.
	\eeq
	The natural operations of the last three types are related by certain equations known as \emph{Fierz identities}~\citep{nishi}. There are other ternary operations which now involve differentiation, such as the two given by
	\begin{align}
		\begin{split}
			\label{spinor_more_ternary}
			(\psi, \phi, \chi) & \longmapsto (\overline{\psi} \sigma^\mu \phi) \tinyspace \partial_\mu \chi,	\\
			(\psi, \phi, \chi) & \longmapsto (\partial^\mu \overline{\psi}) \tinyspace (\partial_\mu \phi) \tinyspace \chi.
		\end{split}
	\end{align}
	To our knowledge, a complete classification of natural operations again remains to be found.
\end{spinor_ex}

In situations where several fields play a role at the same time, one can consider natural operations of mixed arity, by which we mean e.g.~transformations with components of the form
\[
	\Phi_M : \ff_2(M) \times \ff_1(M) \longrightarrow \ff_1(M),
\]
satisfying the two-sorted or more generally multi-sorted analogue of \eqref{naturality}. These arise in particular in the following example.

\begin{spinor_electro_ex}
	\label{spinor_electro2}
	Consider now Weyl spinor electrodynamics as in \Cref{spinor_electro1}, with $\ST$ the category of Lorentzian manifolds and metric-preserving maps.

	Again, first consider the case without gauge invariance. Since the maps are restricted further (relative to \Cref{pure_electro2}) from conformal to metric-preserving, we already obtain a substantial number of additional natural operations on the electromagnetic field functor $\Omega^1$ alone.
	These are defined in terms of the metric and encode geometric information. Among them are the two unary operations given by
	\begin{align}
		\begin{split}
			\label{ddeltad}
			\Omega^1			& \longrightarrow \Omega^1,			\\
			A				& \longmapsto \delta \tinyspace d A,			\\
			A				& \longmapsto d \tinyspace \delta A,
		\end{split}
	\end{align}
	where $\delta := -\star d \star$ is the \emph{codifferential}\footnote{Corresponding to the \emph{divergence} of vector calculus~\citep[Example~4.1.45]{nicolaescu}. Note that, since the Hodge star operator appears twice in it, the codifferential does not depend on an orientation and can likewise be defined on non-orientable manifolds.}. Applying the d'Alembertian is a natural operation given by the sum of these two,
	\beq
		A \longmapsto \Box A = (d \tinyspace \delta + \delta \tinyspace d) A.
	\eeq
	There is also a binary operation given by
	\begin{align}
		\begin{split}
			\label{div_act}
			\Omega^1 \times \Omega^1	& \longrightarrow \Omega^1			\\
			(A,B)				& \longmapsto \left( \delta A \right) \tinyspace B
		\end{split}
	\end{align}
	taking $A$ to a real-valued scalar function $\delta A$, which then acts on $B$ by scalar multiplication. There is a binary operation given by converting two given differential $1$-forms into vector fields via the metric, computing their Lie bracket, and then turning the resulting vector field back into a differential $1$-form. Furthermore, we have the ternary operation given by,
	\begin{align}
		\begin{split}
			\label{Omega1sp1}
			\Omega^1 \times \Omega^1 \times \Omega^1	& \longrightarrow \Omega^1			\\
			(A,B,C)						& \longmapsto ( g^{\mu\nu} A_\mu B_\nu ) \, C_\alpha,
		\end{split}
	\end{align}
	which forms the scalar function given by the inner product of $A$ and $B$ and makes it act by scalar multiplication on $C$. There is a similar ternary operation involving the exterior derivatives of $A$ and $B$, given by
	\begin{align}
		\begin{split}
			\label{Omega1sp2}
			\Omega^1 \times \Omega^1 \times \Omega^1	& \longrightarrow \Omega^1			\\
			(A,B,C)						& \longmapsto g^{\beta\mu} g^{\gamma\nu} (\partial_\beta A_\gamma - \partial_\gamma A_\beta) (\partial_\mu B_\nu - \partial_\nu B_\mu) \, C_\alpha
		\end{split}
	\end{align}
	Note that some of these operations are nonzero only for spacetime dimension $d \ge 2$.

	Concerning mixed natural operations that combine spinors and differential $1$-forms, we can for example combine two spinors into a vector field, and use the metric to turn the result into a differential $1$-form,
	\beq
		\label{mixed_op1}
		(\psi, \phi) \longmapsto g_{\alpha\beta} \tinyspace ( \overline{\psi} \sigma^\beta \phi ).
	\eeq
	We also have a natural operation taking a spinor and a differential $1$-form, combining the two as follows to a new spinor,
	\beq
		\label{mixed_op2}
		(\psi, A) \longmapsto \sigma^\mu A_\mu \psi.
	\eeq
	Taking a suitable linear combination of this natural operation with the Dirac operator results in the gauged Dirac operator, which is itself a natural operation taking a spinor and a differential $1$-form and producing a new spinor,
	\beq
		\label{gauged_dirac}
		(\psi, A) \longmapsto \sigma^\mu (\partial_\mu - i A_\mu) \psi.
	\eeq

As in many other cases, obtaining an understanding---even a conjectural one---of all natural operations seems to be a daring endeavor. We refrain from speculating on whether the above natural operations are sufficient to generate all other ones.

	Upon taking gauge invariance into account, our framework forces us to describe both fields together via a single field functor as in~\Cref{gauge_joint}. This changes the natural operations drastically. For example, we can no longer add two field configurations represented by $(A, \psi)$ and $(B, \phi)$, since the gauge equivalence class of the result $(A + B, \psi + \phi)$ depends on the choice of representatives. But we \emph{can} add the first two components if we replace the spinor component by the zero spinor,
	\[
		\left( (A, \psi), (B, \phi) \right) \longmapsto (A + B, 0),
	\]
	since the gauge equivalence class of the result now only depends on the gauge equivalence classes of the pairs $(A,\psi)$ and $(B,\phi)$. Of course, we similarly also have multiplication by any fixed real number in the spinor component as a natural operation.

	Among the other natural operations that survive gauging is a unary operation given by applying the gauged Dirac operator to the spinor component,
	\[
		(A, \tinyspace \psi) \longmapsto (A, \tinyspace \sigma^\mu (\partial_\mu - i A_\mu) \psi).
	\]
	Another one is the binary operation given by
	\[
		\left( (A, \psi), (B, \phi) \right) \longmapsto \left( B, (\overline{\psi} \psi) \tinyspace \phi \right),
	\]
	which roughly corresponds to~\eqref{spinor_ip} from the pure spinor case, and the gauge invariance is easily verified. There is a similar analogue of~\eqref{fierz1}, given by the ternary operation
	\[
		\left( (A, \psi), (B, \phi), (C, \chi) \right) \longmapsto \left( C, (\overline{\psi} \sigma^\mu \psi) \tinyspace (\overline{\phi} \sigma_\mu \phi) \tinyspace \chi \right),
	\]
	and similarly for~\eqref{fierz2}, which we do not spell out explicitly.

	All of the natural operations given arise from the ungauged case by noting that they are gauge invariant. It is possible that there are additional natural operations in this context which do not arise like this, but we have not yet found any.
\end{spinor_electro_ex}

Let us now return to our general question \ref{which_algebraic2} on the algebraic structure carried by a physical field. Our claim now is that we obtain a good candidate description of this algebraic structure by considering all the natural operations carried by the field functor $\ff : \ST^\op \to \Set$ and interpreting them as algebraic operations.

To explain what this means, we first note that there is a general class of natural operations which are of relevance to the theory itself, and which we have intentionally left out of the discussion of the examples since they are rather general and only of formal abstract interest.
Namely for every $i=1,\ldots,n$, there is a distinguished natural operation which associates to every $n$-tuple of field configurations $(\phi_1,\ldots,\phi_n) \in \ff(M)^{\times n}$ its $i$-th component $\phi_i$. We denote this distinguished natural operation by $\pi_i$, and its component at $M$ by $\pi_{i,M}$.

\newcommand{\FieldLawvere}[2]{\mathrm{Nat}_{#1}({#2})}

So let us denote the set of natural operations of arity $n$ on $\ff$ by $\FieldLawvere{n}{\ff}$. We now explain how this set itself carries an algebraic structure.\footnote{This algebraic structure is known as a \emph{Lawvere theory}~\citep{hyland_power}.} The idea is that natural operations can be composed: if $\Phi \in \FieldLawvere{n}{F}$ and we have arities $k_i$ and natural operations $\Psi_i \in \FieldLawvere{k_i}{F}$ for $i=1,\ldots,n$, then for every manifold $M \in \ST$ we can form the composite map
\[
	\begin{tikzcd}
		\ff(M)^{\sum_{i=1}^n k_i} \ar{rr}{\left(\Psi_1,\ldots,\Psi_n\right)}	&&	\ff(M)^n \ar{rr}{\Phi}	&& \ff(M).
	\end{tikzcd}
\]
This defines a new natural operation of arity $\sum_{i=1}^n k_i$. We denote it by $\Phi \circ (\Psi_1,\ldots,\Psi_n)$.

We now explain how natural operations equip, for every $M \in \ST$, the set of field configurations $\ff(M)$ with algebraic structure.
By definition, every $n$-ary natural operation $\Phi \in \FieldLawvere{n}{\ff}$ implements a map
\[
	\begin{tikzcd}
		\ff(M)^{\times n} \ar{rr}{\Phi_M}		&& \ff(M).
	\end{tikzcd}
\]
The idea now is to consider this map as an algebraic operation on $\ff(M)$ of arity $n$. Moreover, this algebraic structure is such that it respects the composition $\Phi \circ (\Psi_1,\ldots,\Psi_n)$ from the previous paragraph---this holds by definition of the latter. Moreover, the way that the above distinguished natural operation $\pi_i$ acts for $i=1,\ldots,n$ is simply as an actual projection,
\[
	\begin{tikzcd}
		\pi_{i,M} \: : \: \ff(M)^{\times n} \longrightarrow \ff(M), \qquad	(\phi_1,\ldots,\phi_n) \longmapsto \phi_i.
	\end{tikzcd}
\]
Armed with these preparations, we can now axiomatize these operations as an algebraic structure, which will constitute our answer to the question~\ref{which_algebraic2} on the algebraic structure carried by a physical field.

\begin{defn}
	\label{Falgebra}
	Let $\ff : \ST^\op \to \Set$ be a field functor. An \emph{$\ff$-algebra} $\fa$ is a set together with a map
	\[
		a_\Phi \: : \: \fa^{\times n} \longrightarrow \fa
	\]
	for every $n$-ary natural operation $\Phi \in \FieldLawvere{n}{\ff}$, such that the following conditions hold:
	\begin{enumerate}[label=(\roman*)]
		\item Compatibility with composition of operations: for every $\Psi_i \in \FieldLawvere{k_i}{\ff}$,
			\[
				a_{\Phi \circ (\Psi_1,\ldots,\Psi_n)} = a_\Phi \circ (a_{\Psi_1}, \ldots, a_{\Psi_n}).
			\]
		\item Preservation of projections: for every $i=1,\ldots,n$ and $x_1,\ldots,x_n \in \fa$,
			\[
				a_{\pi_i}(x_1,\ldots,x_n) = x_i.
			\]
	\end{enumerate}
\end{defn}

The first equation means more explicitly that for every family of elements $(x_i^j)_{i=1,\ldots,n}^{j=1,\ldots,k_i}$ in $\fa$, we must have
\begin{align*}
	a_{\Phi \circ (\Psi_1, \ldots, \Psi_n)}	& (x_1^1, \ldots, x_1^{k_1}, \ldots, x_n^1, \ldots, x_n^{k_n}) 			\\
						& = a_\Phi \left( a_{\Psi_1}(x_1^1, \ldots, x_1^{k_1}), \ldots, a_{\Psi_n}(x_n^1, \ldots, x_n^{k_n}) \right),
\end{align*}
or in words: $a_{\Phi \circ (\Psi_1, \ldots, \Psi_n)}$, which describes the action of the composite operation $\Phi \circ (\Psi_1, \ldots, \Psi_n)$, must indeed act as the composite of the operation $a_\Phi$ with the $a_{\Psi_i}$.

By the discussion above, the set of field configurations $\ff(M)$ for any manifold $M \in \ST$ is an $\ff$-algebra in a canonical way.
Moreover, the definition of $\ff$-algebra in terms of natural operations guarantees that this is the richest algebraic structure that can be defined on every $\ff(M)$ in a way that respects the symmetries of the field.
Therefore $\ff$-algebras are our answer to  question~\ref{which_algebraic2}.

We now spell out the definition of $\ff$-algebra more explicitly for the particular types of fields considered previously.
Finding a concrete description of $\ff$-algebras in general can be challenging, depending on what field is being considered, since the natural operations are often tricky to classify.
Despite such challenges, these more concrete descriptions will be our starting point in \Cref{sec_physics} for the consideration of the physics of these fields in the absence of a spacetime manifold.

\begin{scalar_fun_ex}
	If we apply \Cref{Falgebra} to  \Cref{scalar_fun1}, then we obtain precisely the definition of $C^\infty$-ring \citep{MR}, with the $n$-ary operations given by the smooth functions $\R^n \to \R$, or equivalently smooth functions of $n$ variables.
\end{scalar_fun_ex}

\begin{scalar_ex}
	\label{pure_scalar3}
	For massless $\phi^4$-theory, we have the rich collection of natural operations partially described in \Cref{pure_scalar2}. Since these natural operations include those of the form~\eqref{Cinfty_natural}, it follows that every $\ff$-algebra is in particular a $C^\infty$-ring.
	
	The additional natural operations which involve the metric now result in a much richer algebraic structure which encodes geometric information on top of the $C^\infty$-ring structure. We thus call $\ff$-algebras of this type \emph{metric $C^\infty$-rings}.
	
	As per \Cref{pure_scalar2}, a metric $C^\infty$-ring $\fa$ in particular comes equipped with a commutative binary operation\footnote{Note that, perhaps in contrast to what this notation suggests, this operation is primitive in our current context.}
	\[
	\langle d -, d - \rangle : \fa \times \fa \longrightarrow \fa,
	\]
	and with a unary operation $\Box : \fa \to \fa$.
	These operations must satisfy equations modelled after those that hold in the manifold case: $\langle d-, d-\rangle$ must be commutative and satisfy the Leibniz rule in the first argument (and by commutativity also the second),
	\beq
		\label{leibniz1}
		\langle d (ab), dc \rangle = a \langle db, dc\rangle + b \langle da, dc\rangle \qquad \forall a,b,c \in \fa,
	\eeq
	as well as the second-order Leibniz rule must hold in the form\footnote{In fact, these equations together show that $\langle d-,d-\rangle$ is uniquely determined by $\Box$, and $\Box$ must be a second-order derivation in the sense of~\citet{GKV}.}
	\beq
		\label{leibniz2}
		\Box(a b) = \Box(a) \hspace{1pt} b + 2 \langle d a, d b \rangle + a \hspace{1pt} \Box (b) \qquad \forall a, b \in \fa.
	\eeq
	The curvature scalar from \Cref{pure_scalar2} will also amount to an operation of arity zero, meaning a fixed distinguished element $R \in \fa$. This element will need to satisfy suitable identities for which we currently do not have an explicit description. It is possible that metric $C^\infty$-rings also carry additional structure or satisfy additional equations which we have not yet found. So due to a lack of understanding of the natural operations involved and the equations between them, we currently do not have a complete and explicit description of metric $C^\infty$-rings. We nevertheless have a range of concrete examples: the very definition of metric $C^\infty$-ring is built such that for every Lorentzian manifold $M$, the set $C^\infty(M)$ is a metric $C^\infty$-ring in a canonical way.
	
	It may be worth noting that having an algebra $\fa$, together with a differential operator acting on it and equipped with an inner product as above, is vaguely reminiscent of the \emph{spectral triples} of noncommutative geometry~\citep{connes,marcolli}.
\end{scalar_ex}

\begin{electro_ex}
	\label{pure_electro3}
	For pure electrodynamics in $d = 4$ dimensions, we saw in \Cref{pure_electro2} that taking linear combinations of the potentials defines a family of natural operations, and we conjectured that there are no other ones, regardless of whether gauge invariance is taken into account or not. If this conjecture is correct, then it follows that the resulting $\ff$-algebras are simply vector spaces, without any additional structure beyond that.
\end{electro_ex}

\begin{spinor_ex}
	\label{pure_spinor3}
	Continuing on with Weyl spinors in four spacetime dimensions from \Cref{pure_spinor2}, the $\ff$-algebras $\fa$ in this case carry a rich algebraic structure as follows. The fact that taking complex linear combinations defines a family of natural operations means that such $\fa$ is in particular a complex vector space. There must be a linear unary operation $D : \fa \to \fa$ corresponding to the Dirac operator.
	
	Furthermore, there must be a host of operations of arity three and five, corresponding to the natural operations~\eqref{spinor_ip}--\eqref{spinor_more_ternary}. Each one of these must be multilinear, since so are the original natural operations in the manifold case. They must also satisfy a bunch of equations amounting to the satisfaction of the Fierz identities. But as already in the scalar field case, we are quite far from having a complete and explicit description of these $\ff$-algebras. We can at least note that spinor field configurations on a Lorentzian manifold are among the examples.
	
\end{spinor_ex}

\newcommand{\electromagnetic}{\mathrm{em}}
\newcommand{\weyl}{\mathrm{W}}

\begin{spinor_electro_ex}
	\label{spinor_electro3}
	The example of Weyl spinor electrodynamics in four spacetime dimensions (\Cref{spinor_electro2}) is clearly our richest example with the largest supply of natural operations, and correspondingly leads to the most complicated type of algebraic structure.
	
	Without taking gauge invariance into account, we have two separate field functors with natural operations of pure and mixed type, and it is therefore most natural to work with the obvious two-sorted analogue of \Cref{Falgebra}. Thus an instance of this type of field algebra structure has two sorts $\fa_{\electromagnetic}$ and $\fa_{\weyl}$, such that $\fa_{\weyl}$ is a type of algebraic structure as described in the previous \Cref{pure_spinor3} and $\fa_{\electromagnetic}$ is analogously an algebraic structure involving operations associated to those discussed in \Cref{spinor_electro2}, such as the d'Alembertian $\Box : \fa_{\electromagnetic} \to \fa_{\electromagnetic}$. There will also be operations mixing the two sorts, corresponding to (among probably others) the mixed natural operations~\eqref{mixed_op1}--\eqref{mixed_op2}, resulting in additional algebraic operations having the types
	\[
	\fa_{\weyl} \times \fa_{\weyl} \longrightarrow \fa_{\electromagnetic}, \qquad \fa_{\electromagnetic} \times \fa_{\weyl} \longrightarrow \fa_{\weyl}.
	\]
	
	Upon taking gauge invariance into account, we had arrived back at a single field functor in \Cref{spinor_electro2}. Correspondingly, in this case the resulting field algebra will also have only one sort, amounting to a sense in which the electromagnetic field gets unified with the spinor field. Due to the examples that we have already given, the reader will have no difficulty with translating the natural operations described in the second half of \Cref{spinor_electro2} into the corresponding $\ff$-algebra structure, although again a complete explicit description remains elusive.
\end{spinor_electro_ex}

It is worth noting that the difficulty with finding explicit descriptions of the $\ff$-algebra structure is not necessarily an impediment towards working with them mathematically.
Indeed the powerful machinery of categorical algebra which applies regardless and provides tools for analyzing algebraic objects even when an explicit description of their algebraic structure is not available~\citep{borceux}. For example, this toolbox allows us to take $\ff$-algebras built out of manifolds and then to perform algebraic constructions on them, such as the formation of quotients. This can result in particular $\ff$-algebras that do not arise from manifolds, but may nevertheless permit a discussion of the physics of the field or fields under consideration, as in the next section.

\section{Algebraic structure relevant for a physical field}
\label{sec_physics}

Finally, we now turn to discussing question~\ref{which_physics2} on how to formulate the physics of a field in purely algebraic terms. By considering general instances of these algebraic structures, we can then lift the physics away entirely from the manifold context.
We also consider how to prune the algebraic structure further in a way which still allows for a formulation of the physics, and outline a few instances of these structures in which the physics still makes sense despite there not being any underlying spacetime manifold.
On the other hand, it is fair to say that we do currently not have a complete satisfactory answer to~\ref{which_physics2}: our treatment of physics is limited to setting up the field equations, and we neither have a satisfactory theory of their solutions, nor do we have a field-algebraic treatment of Lagrangians and the principle of stationary action. Extending field algebra so as to incorporate these aspects of physics remains an open problem.

Based on the previous section, it is natural to try and express the physics of a particular field by using $\ff$-algebras as the relevant algebraic structure.
In doing so, one interprets the elements of an $\ff$-algebra as field configurations, which now make sense without any reference to an underlying spacetime.
Expressing the physics then means in particular to formulate the field equations in purely algebraic terms.

However, the algebraic structures describing physics should be motivated primarily by physical considerations, while the $\ff$-algebras from the previous section are mere mathematical contrivances to begin with.
For the purposes of physics, it may therefore be reasonable to relax these algebraic structures a bit to other ones which can be described concretely, such as in Geroch's proposal based on commutative rings (as opposed to $C^\infty$-rings). 
But the consideration of the $\ff$-algebras can be an important step in practice, based on the idea that the extraction of a suitable algebraic structure from physical considerations can be strongly informed by an understanding of the associated $\ff$-algebras.

So our method is basically to use the $\ff$-algebras as a starting point, and to cut down the natural operations and the $\ff$-algebra structure which they define as physical considerations suggest. We illustrate this method with our running examples.

\begin{scalar_ex}
	To formulate the physics of massless scalar $\phi^4$-theory in $d$ dimensions without reference to a spacetime manifold, 
	we use the metric $C^\infty$-rings $\fa$ from \Cref{pure_scalar3} as a starting point, which are $C^\infty$-rings equipped with additional structure in the form of an additional commutative binary operation $\langle d-,d-\rangle$ and a unary operation $\Box$, satisfying in particular the second-order Leibniz rule, as well as further algebraic operations as per the discussion above.
	We think of the elements of $\fa$ as the possible scalar field configurations, where now an underlying spacetime no longer needs to exist (in whatever form).
	
	Now the key point for physics is that the field equation of massless $\phi^4$-theory---which was our starting point in \Cref{pure_scalar1}---can be expressed purely algebraically on every metric $C^\infty$-ring, where it takes the form
	\beq
		\label{EL}
		\Box \hspace{1pt} \phi = \frac{\lambda}{3!} \phi^3.
	\eeq
	Here, the d'Alembertian $\Box$ is primitive in the algebraic structure on $\fa$, rather than being defined in terms of a Lorentzian metric on a manifold (neither of which makes sense for a general metric $C^\infty$-ring $\fa$).

	Now for the consideration of physics such as the field equation, it seems apparent that not all of the metric $C^\infty$-ring structure is needed. In fact, it may be desirable to now make a move similar to Geroch's: we can define a new algebraic structure modelled after metric $C^\infty$-rings and in which the field equation still makes sense, but which is defined in terms of only finitely many algebraic operations and equations. This can indeed now be done: let us say that a \emph{metric algebra} is a commutative $\R$-algebra $A$ together with
	\begin{itemize}
		\item A commutative binary operation $\langle d -, d -\rangle : A \times A \to A$,
		\item A unary operation $\Box : A \to A$,
	\end{itemize}
	such that both of these are linear (in each argument), and such that the Leibniz rule holds in the form~\eqref{leibniz1} and~\eqref{leibniz2}. Then the field equation~\eqref{EL} still makes sense for every value of the coupling constant $\lambda$.

	This simplified definition seems more prudent in various ways.\footnote{There is one interesting counterpoint to this which shows that also the $C^\infty$-operations may be of some use: they can come in handy in constructing the usual plane wave solutions of the wave equation, since this involves the formation of exponentials (or trigonometric functions), which is part of the $C^\infty$-ring structure but not of the ring structure.}
	One way is that the full $C^\infty$-ring structure builds on the concept of smooth function, the physical relevance of which (at the fundamental level) can be debated.
	Another is that its more explicit nature allows for a simpler mathematical treatment, and in particular allows one to write down and investigate concrete examples quite easily.

	Indeed consider the polynomial ring $\fa := \R[\phi]$, whose elements are polynomials in the field $\phi$, which is now considered as a formal variable. If we define the operations $\langle d-, d-\rangle$ and $\Box$ on the basis elements $(\phi^n)_{n \in \N}$ as
	\[
		\langle d(\phi^n), d(\phi^m) \rangle \,:=\, \frac{\lambda}{12} nm \phi^{n+m+2}, \qquad \Box(\phi^n) \,:=\, \frac{\lambda}{12} n(n+1) \phi^{n+2},
	\]
	and extend linearly, then the relevant equations in the form of the Leibniz rules~\eqref{leibniz1} and~\eqref{leibniz2} are easily seen to hold, so that we indeed have a metric ring in our sense.
	The ring element $\phi \in \fa$ itself now satisfies the field equation~\eqref{EL} by construction, as one can see by taking $n = 1$ in the definition of $\Box$. This example stems from purely technical considerations, and we leave its physical significance open for future work. 
\end{scalar_ex}

\begin{electro_ex}
	The case of pure electrodynamics does not seem to be an interesting example for our field algebra, assuming that the resulting $\ff$-algebras are indeed merely vector spaces (as suggested in \Cref{pure_electro3}). If this speculation is indeed correct, then also the field equations---which are the vacuum Maxwell equations---cannot be expressed in purely field-algebraic terms.

	While this may look like an argument against our proposal, we do not view it in this way because the suggested impossibility of doing so for pure electrodynamics may just reflect the fact that the physics of a universe containing pure electrodynamics as the only physical field would be rather impoverished.  This also shows that the possibility of expressing field equations is not a given, and this underlines that the possibility of doing so in our other examples is a nontrivial feature. 

	This interestingly contrasts with the fact that purely algebraic versions of electrodynamics have been proposed before. One is due to \citet{wise}, where the consideration of higher $p$-form analogues of electrodynamics has in algebraic structures building on those considered in homological algebra (in the form of chain complexes). Similar comments apply to the algebraic gauge theory developed by \citet{zahariev}. In contrast to our field algebra, both of these approaches have non-minimal ontologies in the sense of containing primitive structures that are not physical fields.
\end{electro_ex}

\begin{spinor_ex}
	We treat the Weyl spinor case more concisely, since it is closely parallel to the scalar field case. As per \Cref{pure_spinor3}, an $\ff$-algebra $\fa$ in this case is a complex vector space together with a unary operation $D : \fa \to \fa$ called the \emph{Dirac operator} and additional algebraic operations of higher arity. Thinking of $\fa$ as the set of spinor field configurations without an underlying spacetime, the algebraic structure is obviously rich enough to make sense of the Weyl equation
	\[
		D \psi \,=\, 0,	
	\]
	which is the field equation for a Weyl spinor.

	It now seems like an interesting task to prune the $\ff$-algebra structure, in order to try and come up with an interesting definition of abstract spinor field, parallel to the definition of metric ring in the scalar case. It seems natural to base this around the Dirac operator and the additional operations involved in the Fierz identities discussed before, but we have not yet arrived at a definite proposal for this.
\end{spinor_ex}

\begin{spinor_electro_ex}
	Let us briefly return to Weyl spinor electrodynamics (without gauge invariance). Then as per \Cref{spinor_electro2}, the resulting $\fa$-algebras involve two sorts $\fa_{\electromagnetic}$ and $\fa_{\weyl}$, corresponding to the sets of electromagnetic and the Weyl spinor field configurations, respectively. The Weyl spinor part $\fa_{\weyl}$ is the same as in the previous example, while $\fa_{\electromagnetic}$ now carries a lot of additional algebraic structure as discussed before. Per~\eqref{gauged_dirac}, this structure permits us to express the field equation for the spinor,
	\[
		D_{\mathrm{\scriptscriptstyle{gauged}}} \psi \,=\, 0,	
	\]
	as well as field equations for the electromagnetic field: these are Maxwell's equations with the spinor field as a source term,
	\[
		\delta d A_\alpha \,=\, i g_{\alpha \beta} \bar{\psi} \sigma^\beta \psi,
	\]
	which can be expressed thanks to the natural operations~\eqref{ddeltad} and~\eqref{mixed_op2}. We expect no conceptual differences for these two fields relative to the scalar field and pure Weyl spinor field cases.
\end{spinor_electro_ex}

% physical considerations may hinge on a treatment of the principle of stationary action and the derivation of Euler-Lagrange equations. We do not know how to do this at present, since we have not even entertained the question of how to integrate the Lagrangian density to an action\footnote{It may be worth noting that there are well-known purely algebraic treatments of integration~\citep{segal}, which may come in handy here.}. Here, we merely want to acknowledge these tasks and leave them for future work.

% Similar considerations as in the scalar field case apply to our other examples. As already noted in \Cref{pure_electro2}, a notable exception here is pure electrodynamics in four dimensions, for which we do not seem to obtain any sensible algebraic treatment due to the impoverished nature of the natural operations, ultimately rooted in the conformal invariance of electrodynamics.

To conclude, we would like to highlight the following philosophical features about our formalism. First, it provides a more thoroughgoing dynamic interpretation of relativistic theories than what is available in the literature, such as \citet{brown}. Like Brown's dynamic approach, we treat the metric field on a par with other matter fields rather than representing the primitive geometry of space. When general relativity is not under consideration, we consider metrical information not as a matter field on its own, but as encoded in the algebraic operations of field algebra (as in \Cref{pure_scalar3,pure_spinor3}).  Unlike Brown's approach, we no longer need to posit an underlying metrically amorphous manifold. Physical fields are all we need in our fundamental ontology. 

Second, unlike the other algebraic approaches such as \citet{geroch}, we no longer rely on a scalar field for encoding information about the spacetime structure. As we have argued, such a field need not be among our actual physical fields, and in any case should not be ontologically privileged over other fields. In our approach, we can discern the rich structure of some physical fields and do physics (to some extent) without involving scalar fields at all. All physical fields are put on equal footing. 
	 
Relatedly, in standard algebraicism \`a la Geroch there is no clear distinction between physical fields and mere mathematical fields (various field-like entities are constructed without necessarily representing physical fields). In contrast, we have taken care that the fundamental fields we posited are only those that are standardly acknowledged in physics. For example, we do not posit the Lagrangian of a physical field, which is mathematically represented by a density field, as a physical field itself. Instead, we expect the Lagrangian to be encoded purely in the algebraic structure of the physical field, although we have not implemented this idea yet. While many further tasks await before we can do full-blown physics under our formalism, we consider this work a solid start towards establishing a clean ontology of physical fields without spacetime.

\newpage

\printbibliography

\end{document}